\pdfoutput=1
\documentclass[a4paper,10pt]{article}
\usepackage{acronym}
\usepackage{color}
\usepackage{amsmath,amssymb}
\usepackage{graphicx}
\usepackage{footmisc}
\usepackage{hyperref}
\usepackage{bbold}

\newcommand{\qsg}{q_\mathrm{SG}}
\newcommand{\qcg}{q_\mathrm{CG}}
\graphicspath{{../figuras/}{./}}

\begin{document}

\title{Precursors of the Spin Glass Transition in Three Dimensions}

\author{Marco Baity-Jesi$^{1,\dagger}$, V\'ictor Mart\'in-Mayor$^{2}$\\
\small 1 Department of Chemistry, Columbia University, New York, NY 10027, USA\\
\small 2 Departamento de F\'isica Te\'orica, Universidad Complutense de Madrid, 28040 Madrid, Spain\\
\small $\dagger$ mb4399@columbia.edu}

\date{\today}

\maketitle

\begin{abstract}
We study energy landscape and dynamics of the three-dimensional
Heisenberg Spin Glass model in the paramagnetic phase, i.e. for
temperature $T$ larger than the critical temperature
$T_\mathrm{c}$. The landscape is non-trivially related to the
equilibrium states even in the high-temperature phase, and reveals an
onset of non-trivial behavior at a temperature $T_\mathrm{o}$, which
is also seen through the behavior of the thermoremanent magnetization.
We also find a power-law growth of the relaxation times far from
the spin-glass transition, indicating a dynamical crossover at a
temperature $T_\mathrm{d}$, $T_\mathrm{c}<T_\mathrm{d}<T_\mathrm{o}$.
The arising picture is reminiscent of the phenomenology of supercooled
liquids, and poses questions on which mean-field models can describe
qualitatively well the phenomenology in three dimensions. On the
technical side, local energy minima are found with the Successive
Overrelaxation algorithm, which reveals very efficient for energy
minimization in this kind of models.
\end{abstract}

\section{Introduction}
Spin glasses are disordered magnetic alloys~\cite{mezard:87}.  It is
well established that, in three dimensions, they exhibit a phase
transition from a paramagnetic to a \emph{spin glass} phase at a
critical temperature $T_\mathrm{c}$, where the spins freeze in
amorphous configurations. Evidence for this phase transition is
experimental~\cite{denobel:59,mydosh:93},
numerical~\cite{palassini:99,ballesteros:00,lee:03,boettcher:05} and
(to some extent) also analytical~\cite{parisi:94}.\\

The nature of the low-temperature phase, the spin glass phase, is not
well understood yet.  On one side, contrasting interpretations of
numerical simulations have been
supported~\cite{janus:10,larson:13,janus:14b,janus:14c,janus:17b,wang:17},
and on the other side, first-principles analytical calculations can be
performed only in the mean-field limit, corresponding to infinite
dimensions~\cite{parisi:79,elderfield:82}.

Maybe the simplest scenario for the spin-glass phase is the Droplet
Picture, according to which this phase is not very dissimilar to ferromagnetic
ordering, with only two pure states related by
symmetry~\cite{bray:85,mcmillan:85,fisher:86}. The other dominant
theory is the Replica Symmetry Breaking (RSB) scenario, according to
which three-dimensional spin glasses behave similarly to their
mean-field counterparts~\cite{parisi:96,marinari:00}. Since different
spin-glass models display qualitatively different features, the RSB
scenario can apply in different ways.

The main mean-field model taken into account by the RSB scenario is the SK model~\cite{sherrington:75}, which is conjectured to behave similarly to the three-dimensional Ising spin glass~\cite{janus:08b,janus:10b}.
Another mean-field model with different phenomenology is the $p$-spin model~\cite{derrida:80,crisanti:92,cugliandolo:93,crisanti:04}, which is more often used to interpret results for three-dimensional structural glasses rather than spin glasses.\footnote{Unless specified otherwise, throughout this work we will refer as $p$-spin model both for the pure~\cite{derrida:80} and for the mixed~\cite{crisanti:04} $p$-spin models.}
This is due to striking similarities with the phenomenology of supercooled liquids (SCLs), and to the presence of several non-trivial effects even in the paramagnetic phase~\cite{kirkpatrick:89, cavagna:09, charbonneau:14}.

These effects include the presence of a dynamical temperature $T_\mathrm{d}>T_\mathrm{c}$, associated to a so-called \emph{topological transition}: at temperature $T>T_\mathrm{d}$ the phase space motion of the system is controlled by saddles of the energy landscape, whereas for $T<T_\mathrm{d}$
the typical configurations are close to local minima of the energy. Since the energy barriers in the $p$-spin model are diverging~\cite{cugliandolo:95}, this implies that the system becomes stuck around the local minima, the relaxation times diverge, and ergodicity is broken. Furthermore, instant energy minimizations (quenches) starting from equilibrium configurations at $T\geq T_\mathrm{d}$ converge almost surely to local minima (also called inherent structures) at a fixed \emph{threshold energy} $E_\mathrm{th}=E(T_\mathrm{d})$, and only by equilibrating at lower temperatures it is possible to reach lower energies.

A similar phenomenology was encountered in models of supercooled liquids, marking the onset of glassy behavior. 
On one side, an apparent power-law divergence 
of the relaxation times in the high-temperature phase is present~\cite{kob:94}.
On the other side, at high temperatures the energy of the inherent structures (IS) is constant, until it starts decreasing at an onset temperature $T_\mathrm{o}$~\cite{sastry:98}. 
In low-dimensional systems these transitions are not sharp, because the energy barriers are not infinite, so they are only found in the form of crossovers.

These effects, that occur in the disordered phase, do not occur in the  the SK model, which, in some sense, has a more trivial high-temperature phase. 
For this reason one usually refers to the $p$-spin model as a mean-field structural glass~\cite{charbonneau:14} and to the SK as a mean-field spin glass.
\\

Here, we show that a structural-glass-like (or $p$-spin-like) phenomenology is present
also in three-dimensional spin glasses of the Heisenberg type.
Specifically, we find that, as it is observed in supercooled liquids,
the IS energy has a non-trivial dependence on temperatures, and that
the relaxation times have a power-law growth deep in the paramagnetic
phase, which can be interpreted as a dynamical crossover.  Also, we
find that the memory of the initial condition after a quench resembles
the $p$-spin rather than the SK model.\\

We introduce the model and the main observables in
Sec.~\ref{sec:model}.  In section~\ref{sec:sor} we explain how
inherent structures can be easily found by using a simple local
algorithm. The main results of the paper are given in
section~\ref{sec:results}, followed by a discussion in
section~\ref{sec:conc}.

\section{Model and Observables}\label{sec:model}
We study the Heisenberg spin glass model on a three-dimensional periodic cubic lattice of linear size $L$. The system has $N=L^3$ spins
\footnote{
We study $L=64, 128$, finding consistent results, which indicates that our analysis is not hampered by finite-size effects. Therefore, in our analysis, we assume that the system is in the thermodynamic limit.
}
$\vec{s}_i=(s_{i,x},s_{i,y},s_{i,z})$, with $\vec{s}_i^{\,2}=1$. The Hamiltonian is
\begin{equation}\label{eq:H}
 \mathcal{H} = -\frac13\sum_{\langle i,j\rangle} J_{ij} \vec s_i\cdot \vec s_j\,,
\end{equation}
where $\langle i,j\rangle$ indicates that the sum is performed over nearest neighbors, and the couplings $J_{ij}$ are randomly chosen from a Gaussian distribution with mean zero and variance one.
For averages over the couplings (disorder averages), we use an overline, $\overline{(\ldots)}$, whereas for averages over the Gibbs distribution (thermal averages) we use angle brackets, $\langle\ldots\rangle$.

\bigskip
This model has a phase transition from a paramagnetic to a glassy phase at $T_\mathrm{c}=0.120(1)$~\cite{fernandez:09b}.
There are two order parameters: the spin-glass overlap $\qsg$ and the chiral-glass overlap $\qcg$. 
Some numerical works indicate two separate transitions temperatures spin and chirality, with the chirality at a higher temperature~\cite{viet:09}, while others assess no measurable difference between the two~\cite{fernandez:09b}. To the scope of this work, the only important datum is that temperatures higher than $T_\mathrm{c}$ are in the paramagnetic phase.

In order to have a rotation-invariant definition of the overlap, we use the squared overlap $\qsg^2$, defined through
\begin{align}
 \tau_{i,\alpha\beta} =& s_{i,\alpha}^{(a)}s_{i,\beta}^{(b)}\\
 Q_{\alpha\beta}      =& \sum_{i=1}^{N} \tau_{i,\alpha\beta} \\
 \qsg^2               =& \frac1{N^2}\mathrm{Tr(Q^\dagger Q)} & \label{eq:qsg}
\end{align}
where $\alpha, \beta=x,y,z$ are the components of the spins, $(a), (b)$ indicate two different configurations with the same realization of the couplings $\{J_{ij}\}$ (replicas).

For the chiral overlap, instead, we need to define the chirality along the spatial direction $\mu=x,y,z$,
\begin{equation}
 \kappa^{\mu}_i = s_{i:-\mu}\cdot ( s_{i} \times s_{i:+\mu} )\,,
\end{equation}
where $s_{i:-\mu}$ (or $s_{i:+\mu}$) is the neighbor of $s_{i}$ along the negative (or positive) direction $\mu$.
The chiral overlap between two replicas $(a)$ and $(b)$ in the direction $\mu$ is then defined as
\begin{align}\label{eq:qcgmu}
 \qcg^\mu =& \frac1N\sum_{i=1}^N \kappa^{\mu,(a)}_i \kappa^{\mu,(b)}_i\,,\\
 \qcg     =& \frac13\sum_{\mu=x,y,z} \qcg^\mu\,.
\end{align}

Correlation functions $C(r)$ (and their Fourier transforms, the wave-number dependent susceptibilities $\chi(k)$) are measured per planes (applying Eqs.~\eqref{eq:qsg} and \eqref{eq:qcgmu} to couple different sites). The correlation along the direction $\hat{e}_1$ is 
\begin{equation}
 C_1(r) = \frac{1}{N}\sum_{x=1}^L P(x)P(x+r)\,
\end{equation}
where $P(x)$ is the average overlap (either spin or chiral glass) along the $x^\mathrm{th}$ plane perpendicular to $\hat{e}_1$.
Given the isotropy of the problem, we averaged over the three main directions: 
\begin{equation} 
C(r) = \frac13\left[C_1(r)+C_2(r)+C_3(r)\right]\,.
\end{equation}
The chiral correlations can have the chirality direction $\mu$ either parallel or orthogonal to axis of the correlation. We do not remark difference behaviors between the two (see~\cite{baityjesi:11}), so we show results averages over all directions.
The susceptibilities $\chi(k)$ were calculated by Fourier transforming $C_{1}(x),C_{2}(x)$ and $C_{3}(x)$, and then averaging.

In order to diminish statistical errors, we follow~\cite{janus:09b},
and truncate the correlation functions when the signal becomes less than three times the error bar (see~\cite{baityjesi:11} for details and finite-size effects).
From those we calculated the correlation length
\begin{equation}
 \xi = \frac{1}{2\sin(\frac\pi L)} \sqrt{\frac{\chi(0)}{\chi(\frac{2\pi}{L})}-1}\,.
\end{equation}

Similarly, we calculate the spin-glass autocorrelation functions
\begin{equation}
 C_\mathrm{SG}(t) = \frac1{N^2}\mathrm{Tr}\left(Q_{0,t}Q_{0,t}^\dagger \right)\,
\end{equation}
where $Q_{0,t}=\sum_i^N s_{i,\alpha}(0) s_{i,\beta}(t)$ is the overlap calculated between configurations of the same replica at times $0$ and $t$.
Analogously, the chiral glass correlation functions can be written as
\begin{equation}
 C_\mathrm{CG}(t) = \frac{1}{3N}\sum_{\mu=x,y,z}\sum_{i=1}^N \kappa^\mu_i(0) \kappa^\mu_i(t)\,.
\end{equation}

The integrated autocorrelation times are 
\begin{align}
\label{eq:autocorrSG} \tau_\mathrm{SG} &= \int_0^\infty C_\mathrm{SG}(t)\,dt\,, \\
\label{eq:autocorrCG}  \tau_\mathrm{CG} &= \int_0^\infty C_\mathrm{CG}(t)\,dt\,.
\end{align}

\section{Inherent Structures}\label{sec:sor}
The most intuitive way to perform a direct minimization on a spin model is to successively align every spin $\vec s_i$ to its local field
\begin{equation}
 \vec h_i = \sum_{\langle i,j\rangle} J_{ij}\vec s_j\,.
\end{equation}
This algorithm, called the Gauss-Seidel (GS) algorithm~\cite{sokal:92}, has an extremely slow convergence rate, and is not efficient on our system sizes~\cite{varga:62}.
\footnote{Though it has been used previously, for example in~\cite{baityjesi:15}, for finding inherent structures in small Heisenberg spin glasses.}
Therefore, we resort to a modification of the GS algorithm, called Successive Overrelation (SOR), that consists essentially in adding momentum to the minimization~\cite{sokal:92}.

On the Heisenberg spin glass, the SOR algorithm amounts to updating the spins sequentially, with a linear combination between the local update that mostly decreases the energy (i.e. the GS update)
\begin{equation}
 \vec s_i^\mathrm{\,GS} = \frac{\vec h_i}{|\vec h_i|}\,,
\end{equation}
and the one that mostly changes the spin position without 
changing its energy
\footnote{This update is the overrelaxation update which is often used for equilibrating Heisenberg spin glasses~\cite{campos:06}.}
\begin{equation}
 \vec s_i^\mathrm{\,OR} = 2\frac{\vec s_i\cdot\vec h_i}{\vec h_i^{\,2}}\vec h_i - \vec s_i\,.
\end{equation}

The SOR update, $\vec s_i^\mathrm{\,SOR}$, 
\begin{equation}\label{eq:sor}
 \vec s_i^\mathrm{\,SOR} = \frac{\vec s_i^\mathrm{\,GS} + \lambda \vec s_i^\mathrm{\,OR}}{|\vec s_i^\mathrm{\,GS} + \lambda \vec s_i^\mathrm{\,OR}|}\,.
\end{equation}
depends on a parameter $\lambda>0$, which can be increased to make the
minimization less greedy. Indeed, increasing $\lambda$ reduces the
greediness the algorithm since with $\lambda=0$ one recovers GS, and
with $\lambda=\infty$ SOR is recovered.

\begin{figure}
\centering
\includegraphics[width=0.8\textwidth]{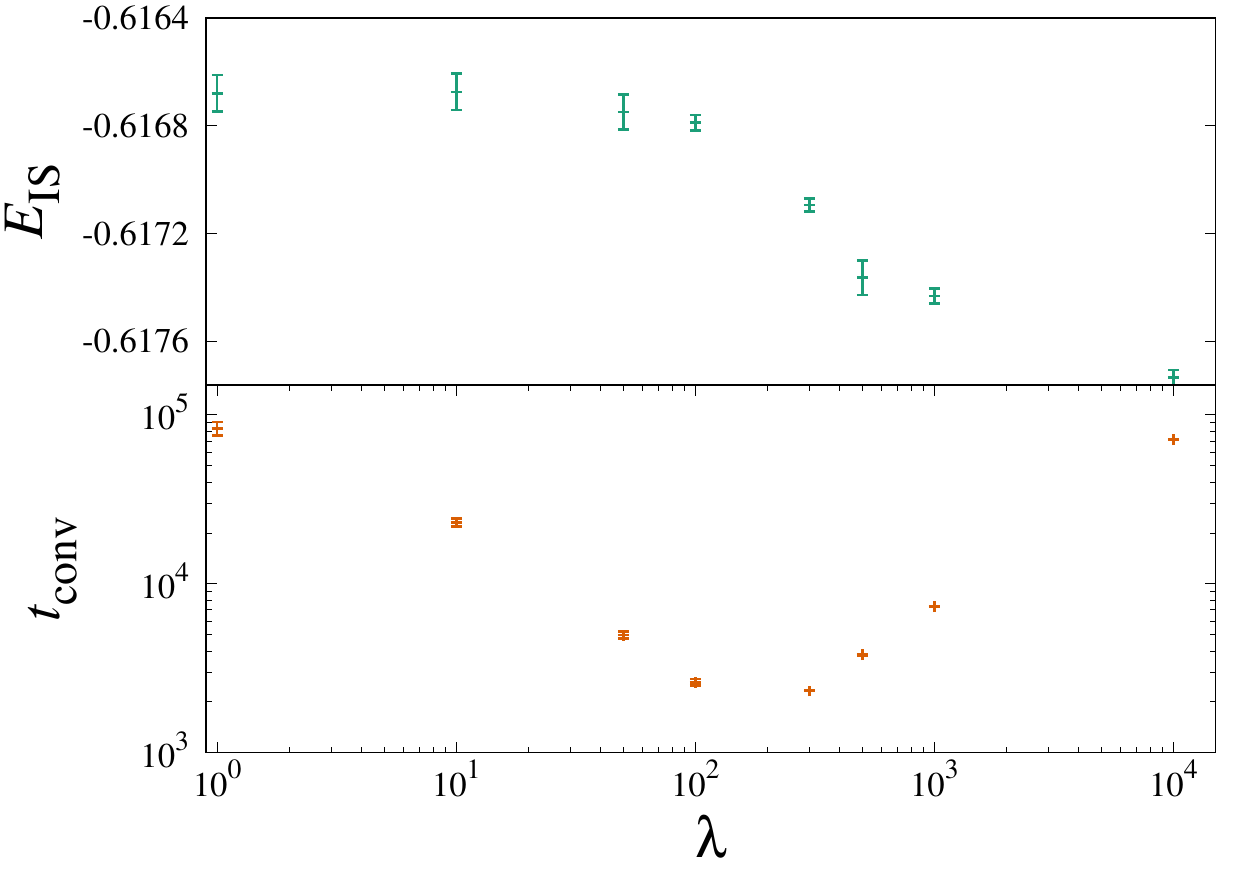} 
\caption{Top: energy of the IS reached with Successive Overrelation, as a function of the parameter $\lambda$ (Eq.~\ref{eq:sor}),
starting from $T=\infty$.
Bottom: convergence time of the SOR algorithm. The minimization was considered converged when the maximum spin displacement in a sweep was smaller than $10^{-14}$.
}
\label{fig:sor}
\end{figure}

Ungreedy versions (i.e. large $\lambda$) of the algorithm are able to
reach lower energies (Fig.~\ref{fig:sor}, top), in agreement with
similar algorithms on Ising spin models~\cite{parisi:03}.

In order to achieve optimal performance, $\lambda$ must be large enough to be away from the GS limit, but not too large, otherwise the energy does not get minimized at all. It turns out that the best performance is obtained in the interval $\lambda\in[100,300]$ (Fig.~\ref{fig:sor},bottom).
\footnote{For more tests and applications of SOR to the Heisenberg spin glass see~\cite{baityjesi:11}.}

\section{Results}\label{sec:results}
We thermalize configurations deep in the paramagnetic phase, at temperatures $T$ = 0.19, 0.21, 0.23, 0.25, 0.3, 0.5, and $\infty$ with a combination of the heatbath and overrelaxation algorithms~\cite{campos:06,lee:07}.

\begin{figure}
\centering
\includegraphics[width=0.8\textwidth]{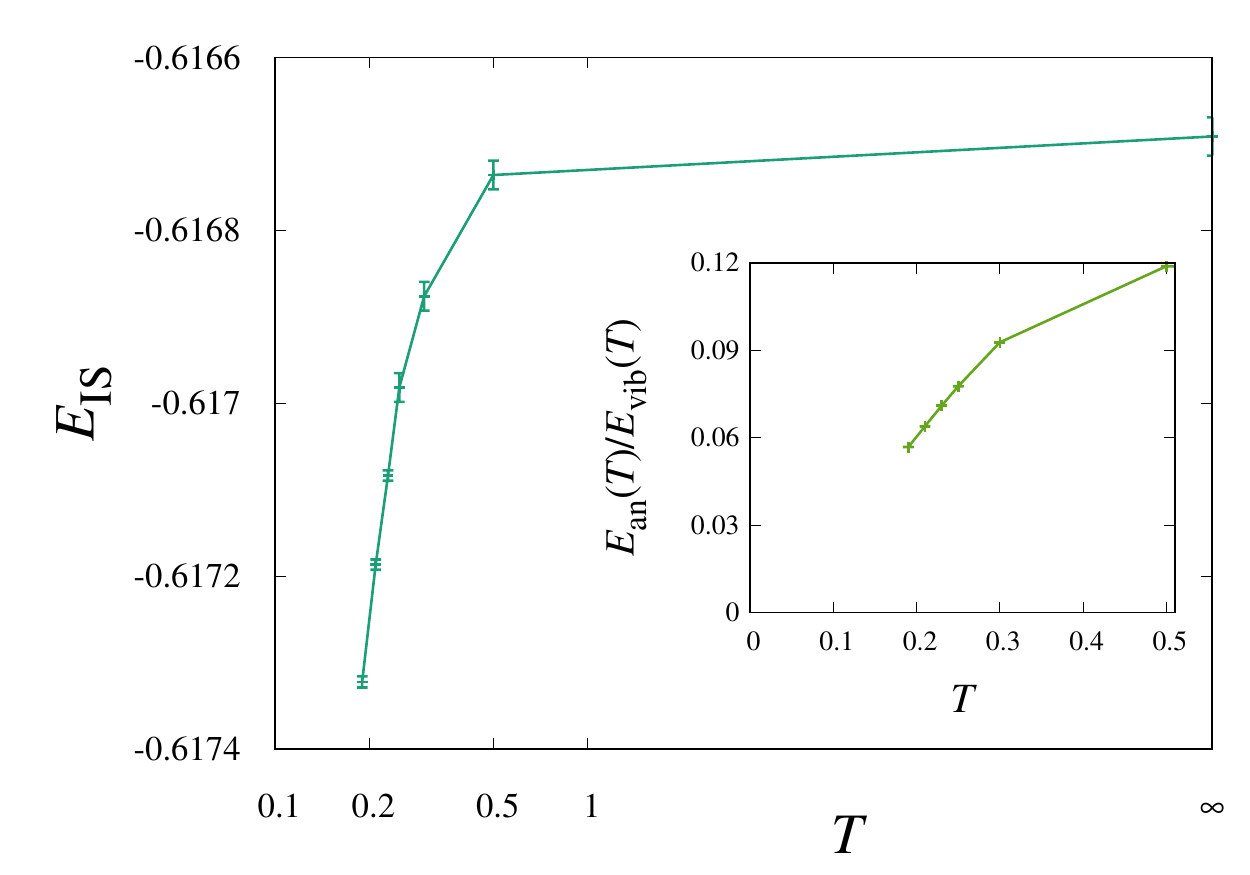}
\caption{
Main figure: energy of IS reached when starting the minimization from equilibrium configurations at temperature $T$.
Inset: ratio between the anharmonic contribution to the energy, $E_\mathrm{an}$, and the vibrational energy $E_\mathrm{vib}$.
}
\label{fig:eis}
\end{figure}

\paragraph{Inherent Structure Energy}
From each configuration, we minimize the energy with SOR, setting $\lambda=300$ (results with $\lambda=100$ reveal equivalent). 
As shown in Fig.~\ref{fig:eis}, the IS energy is weakly dependent on temperature, and decreases abruptly at a temperature consistently higher than the spin-glass transition temperature $T_\mathrm{c}=0.120(1)$~\cite{fernandez:09b}, reminescently of what occurs in supercooled liquids~\cite{sastry:98}, where this is interpreted as evidence that the energy landscape plays a pivotal role in the slowing down of glassy systems. 
Further, the energy of a low-temperature state, $E$, can be decomposed in three terms: the energy of the inherent structure, $E_\mathrm{IS}$; a  vibrational component $E_\mathrm{vib}=k_\mathrm{B}T/3$; 
\footnote{The factor 1/3 in in the vibrational energy stems from the normalization we chose in Eq.~\ref{eq:H}.}
and an anharmonic energy term, $E_\mathrm{an}$, that is zero in the case that the energy consist purelly of harmonic vibrations around the minimum.
From this decomposition, we show in Fig.~\ref{fig:eis}--inset that the anharmonic term $E_\mathrm{an}(T)=E(T)-E_\mathrm{IS}(T)-k_\mathrm{B}T/3$ is a small fraction of the vibrational energy, for which an energy landscape interpretation of the energy of a state as principally an
inherent structure plus harmonic thermal fluctuations is reasonable even in the paramagnetic phase.

\begin{figure}
\centering
\includegraphics[width=0.8\textwidth]{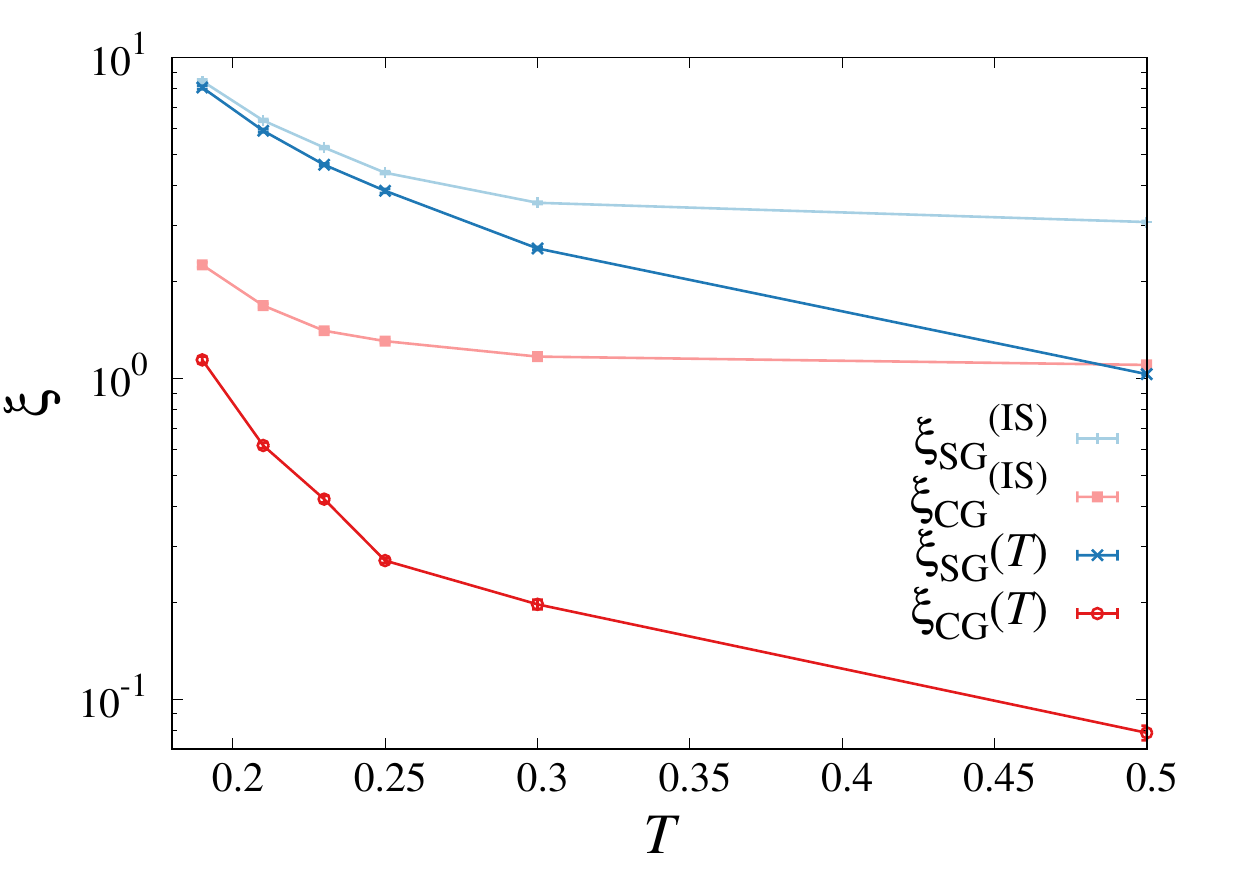} 
\caption{Spin (blue) and chiral glass (red) correlation lengths of the equilibrium state (dark) and of the inherent structure (light).}
\label{fig:xi}
\end{figure}

\paragraph{Length Scales}
As shown in Fig.~\ref{fig:xi}, the qualitative trend of the IS energy is reflected by the IS correlation length, $\xi^\mathrm{(IS)}$, which starts increasing at $T\approx0.3\gg T_\mathrm{c}$.
At higher temperature, $\xi^\mathrm{(IS)}$ is approximately constant, and significantly larger than the equilibrium correlation length $\xi$, that goes to zero as $t\to\infty$. 
At lower temperature, the growth of the correlations becomes progressively dominated by the landscape, especially in the spin-glass sector, where the $\xi$ and $\xi^\mathrm{(IS)}$ almost coincide.
In the chiral sector there is also a growing correlation length, but, at least at higher temperature, it can not be associated to a growth of the landscape correlations. This is similar to what happens in supercooled liquids, with the crucial difference that in the Heisenberg spin glass we know that at a low-enough temperature the correlations will eventually diverge~\cite{fernandez:09b}.

Critical scaling would require that the correlation length diverge at the phase transition as $\xi\sim(T-T_\mathrm{c})^{1/\nu}$, where the critical exponent $\nu=1.5$ can be taken from Ref.~\cite{fernandez:09b}.
The best fit we are able to obtain from gives a divergence of $\xi_\mathrm{SG}$ at $T_\mathrm{SG}=0.095(2)$, which is incompatible with its most accurate measurement, $T_\mathrm{c}=0.120(1)$~\cite{fernandez:09b}.
Equivalently, fixing the critical temperature yields $\nu=1.146(9)$, again incompatible with its value at $T_\mathrm{c}$, so this temperature range the growth of the correlation length is not fully dominated by the phase transition.

\begin{figure}
\centering
\includegraphics[width=0.8\textwidth]{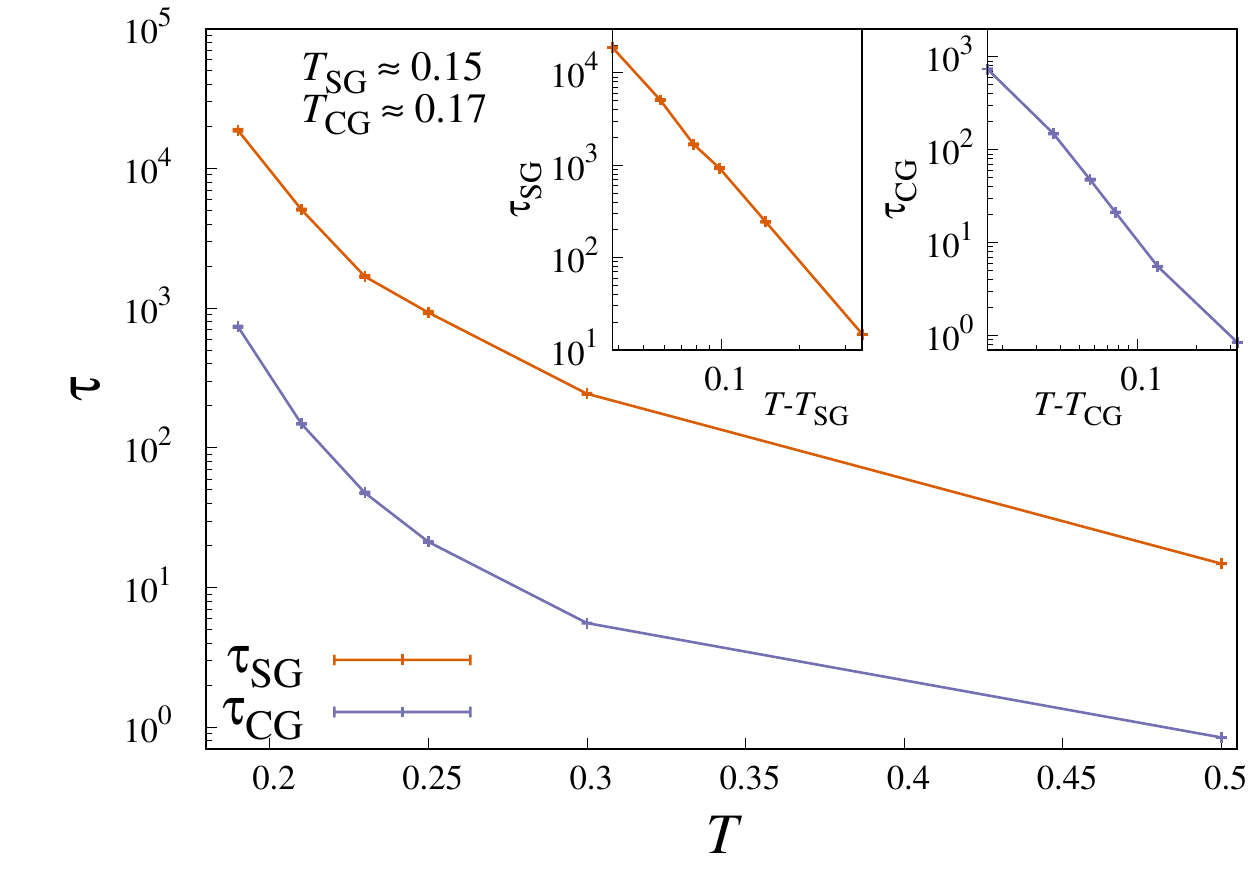} 
\caption{Spin (orange) and chiral glass (blue) relaxation times. Power law fit extrapolations give a divergence of the relaxation times at $T_\mathrm{SG}=0.152(3)$ and $T_\mathrm{CG}=0.173(4)$. The insets show the same data on a different scale, to emphasize the power-law behavior.
Note that these are not real divergences, since we are in the deep paramagnetic phase. $T_\mathrm{SG}$ and $T_\mathrm{CG}$ should be regarded instead as the temperatures around which the activated dynamics starts to be relevant.
}
\label{fig:tau}
\end{figure}

\begin{figure}[tb]
\centering
\includegraphics[width=0.8\textwidth]{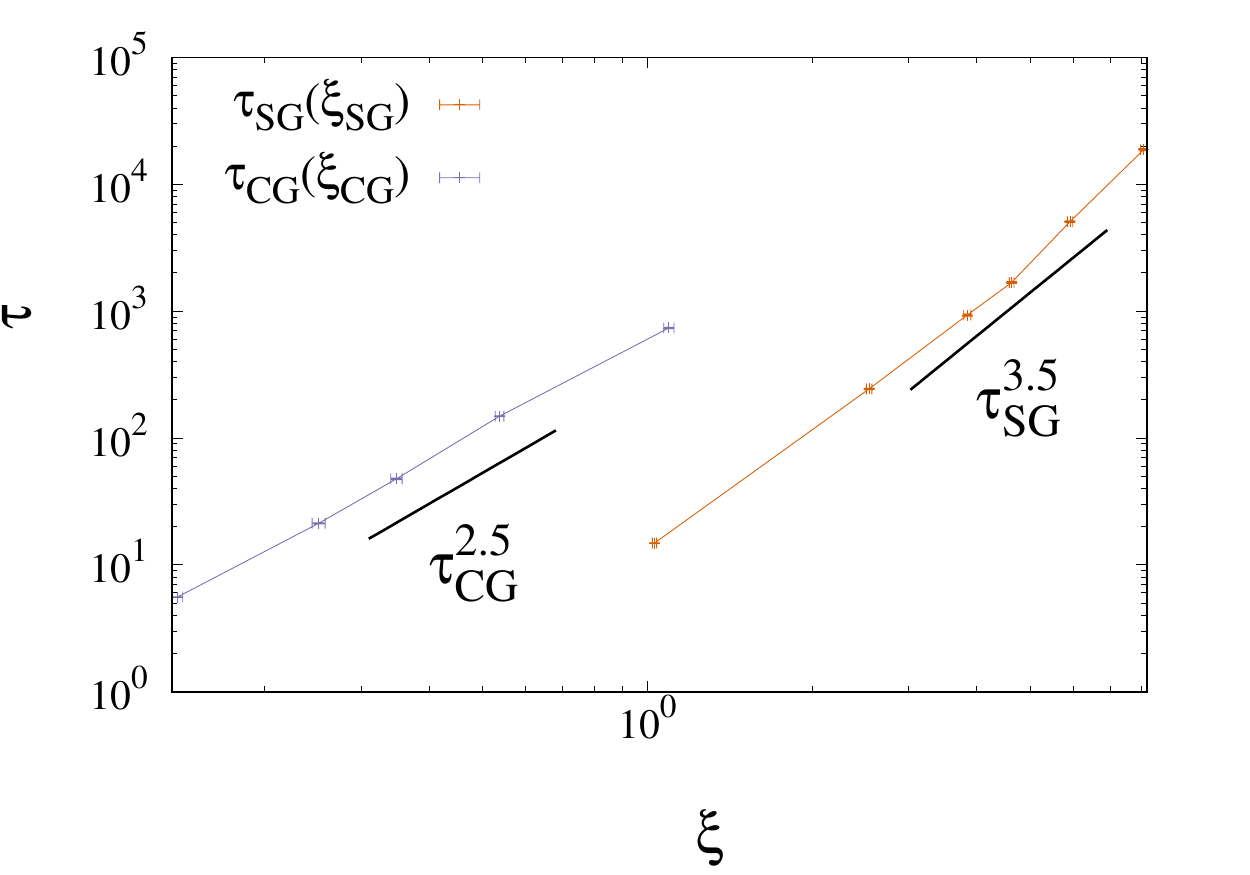} 
\caption{Spin (orange) and chiral glass (blue) relaxation times as a function of the corresponding correlation length. The black lines show some reference power laws.
}
\label{fig:tau-xi}
\end{figure}

\paragraph{Relaxation Times}
Starting from the thermalized configurations, with run pure heatbath dynamics to measure the autocorrelation times $\tau_\mathrm{SG}$ and $\tau_\mathrm{CG}$ (Eqs.~\eqref{eq:autocorrSG} and \eqref{eq:autocorrCG}.
Fig.~\ref{fig:tau} illustrates that, in the same temperature range where $\xi$ grows mildly, the integrated autocorrelation times grow of about three orders of magnitude. In the spin glass sector, the relaxation time grows from 15ps at $T=0.5$ to 19000ps at $T=0.19$. The chiral relaxation time goes from 1ps to 700ps.
\footnote{
One heat-bath sweep in our simulations roughly corresponds to 1ps of physical time~\cite{mydosh:93}.
}
In the insets we show that this growth follows a power law,
reminiscent of the mode-coupling crossover in SCLs~\cite{kob:94}. The
extrapolated transition temperatures in the spin and chiral glass
channels are respectively $T_\mathrm{SG}=0.152(3)$ and
$T_\mathrm{CG}=0.173(4)$.  We stress that these are not real
transitions, but mere crossovers: these power-law growths are expected
to be smoothed down when the temperature is low enough for activated
dynamics to be relevant.  The exponent's value is $3.2(1)$ for the
spin and $2.3(2)$ for the chiral sector.  A similar decoupling between
spin and chirality was also observed in~\cite{picco:05}, through AC
susceptibility measurements in the same temperature range.\footnote{
We know that there can not be a power law divergence at this
temperature, and that this growth necessarily a crossover, implying
that there is a bias error in addition to the statistical error. As a
consequence, for these fits, we use errorbars. The $\chi^2/d.o.f.$ is
4.50/3 (with $6\%$ errors) for the spin glass sector, and 5.98/3 (with
$13\%$ errors) for the chiral glass.}  This growth of $\tau$ is consistently
stronger than the one of the correlation lengths. The relationship
between the two is shown in Fig.~\ref{fig:tau-xi}. It is not clear
what relationship there is between the two observables, because the
correlation length's extrapolation would (wrongly) suggest a
divergence at $T<T_\mathrm{c}$, while the relaxation times' apparent
divergence is at $T>T_\mathrm{c}$.

\paragraph{Thermoremanent magnetization}
Further similarities with a SCL-like behavior (therefore, with the $p$-spin model) are seen in the thermoremanent magnetization, defined as the overlap between the equilibrium configuration and the corresponding inherent structure.\\
\begin{figure}[tbh]
\centering
\includegraphics[width=0.8\textwidth]{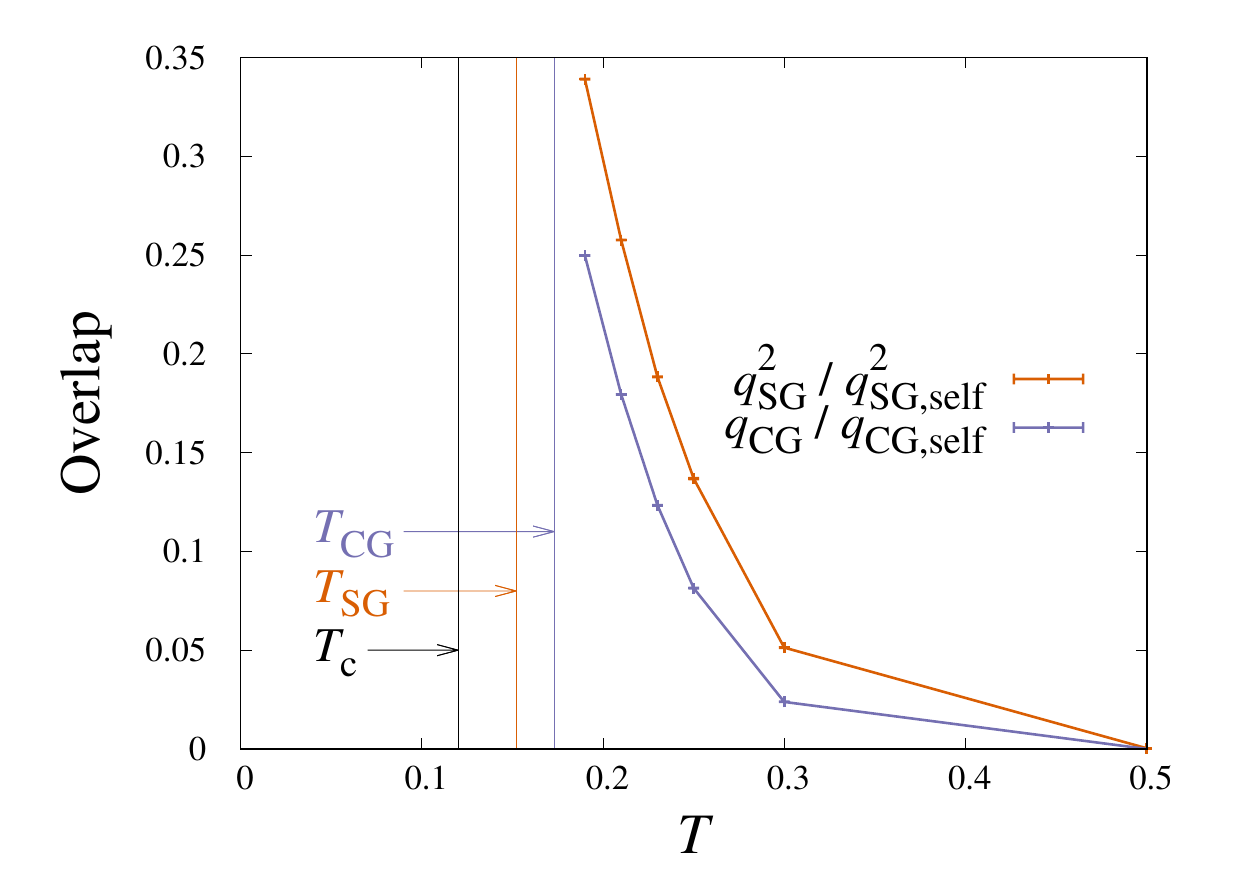}
\caption{
Overlap between the equilibrium configuration and its inherent structure, normalized with the self-overlap.
This is not the equilibrium overlap, it is a non-equilibrium quantity.
The leftmost vertical line shows the thermodynamic spin-glass transition temperature, $T_\mathrm{c}$. The other two vertical lines represent the temperatures at which the extrapolations of $\tau_\mathrm{SG}(T)$ and $\tau_\mathrm{CG}(T)$ diverge (they do not represent a phase transition: they represent the mode-coupling crossover). In the $p$-spin model, the plotted quantity is expected to reach a value $0<q^*<1$ at the dynamic transition.
}
\label{fig:thermoremanent}
\end{figure}
We show in Fig.~\ref{fig:thermoremanent} that it is zero at large $T$, and it becomes larger than zero at temperatures consistently higher than $T_\mathrm{c}=0.120(1)$. 
This reveals a transition (or crossover) from a trivial temperature regime in which memory after a quench is totally lost, to a regime with glassy behavior. The temperature of this onset of glassy behavior is consistent with the one revealed by the inherent structure energy (Fig.~\ref{fig:eis}).\\
This behavior of the thermoremanent magnetization is qualitatively different both from the SK model (where it is non-zero even at infinite temperature~\cite{folena:19}) and
from the pure $p$-spin model (where a finite jump is expected at $T_\mathrm{d}$).
It is instead expected in the mixed $p$-spin model~\cite{riccitersenghi-internal}.

\section{Discussion}\label{sec:conc}
We analyzed the ISs in the three-dimensional Heisenberg spin glass, with the Successive Overrelaxation algorithm, which reveals efficient even for very large system sizes, and is straightforwardly implemented. 
Our findings are consistent with an onset temperature of glassy behavior, $T_\mathrm{o}$, where the IS energy starts depending on temperature and the thermoremanent magnetization becomes non-zero.
This is accompanied by a gentle growth of the correlation lengths $\xi$, both chiral and spin glass, which, at least at higher temperatures, appears unrelated to the landscape. 
While the growth of $\xi$ is mild, the autocorrelation times appear to diverge as a power law to a temperature that is sizably higher than the critical temperature $T_\mathrm{c}$,\footnote{This power-law growth is expected to be leveled down at lower temperatures by activated processes, since we are in the paramagnetic phase.} 
suggesting a mode-coupling transition (which in 3D is a crossover, not a phase transition). 

The aforementioned features are qualitatively similar to what occurs in supercooled liquids (SCLs) and in the $p$-spin model~\cite{berthier:11}.
There is a longstanding debate over whether spin glasses and SCLs display the same critical behavior, due to the formal equality between the dynamical equations of the $p$-spin glass and the mode-coupling equations for SCLs~\cite{kirkpatrick:87,kirkpatrick:89}. This equivalence is well assessed in mean field, both from the point of view of the statics~\cite{charbonneau:14} and the dynamics~\cite{crisanti:00b,maimbourg:16}, and is fortified by recent evidence that the dynamical solution of hard spheres~\cite{maimbourg:16} can be obtained through a mode-coupling approach~\cite{szamel:17}.

In three dimensions, the random first order transition theory~\cite{berthier:11} supports an equivalence between spin glasses and SCLs~\cite{berthier:19}, though there are opposing views~\cite{franz:11,wyart:17}.
Moreover, there is some suggestion that SCLs are in the same universality class of Ising spin glasses in an external magnetic field~\cite{moore:02}, which breaks time-reversal symmetry of the SG, that is absent in SCLs.

To our knowledge, this is the first study that explores this equivalence on spin glasses with continuous degrees of freedom, which may explain why the features we showed had not yet been found in a three-dimensional spin glass.
Our findings strongly suggest that Heisenberg spin glasses behave very similarly to SCLs.\footnote{
Though perturbations to a Heisenberg hamiltonian, such as exchange anisotropy result into an Ising universality class~\cite{baityjesi:14}. 
}
This is further supported by striking similarities in the energy landscape of these models in three dimensions: both Heisenberg spin glasses~\cite{baityjesi:15b} and SCLs~\cite{lerner:16} have an excess of soft localized modes (at frequencies much lower than the boson peak) with the same exponent $g(\omega)\propto\omega^4$. 

The presence of a dynamical transition in three-dimensional SGs supports also a connection with results obtained within mode-coupling theory~\cite{reichman:05}, which could hold in three dimensions~\cite{szamel:17} under hypotheses that we are currently verifying~\cite{baityjesi:19}.

Especially the chiral order parameter shows a behavior reminiscent of SCLs. In fact, it displays a very mild growth of the correlation length in a temperature regime where the relaxation times explode, which is one of the main puzzles in the glass transition problem~\cite{berthier:11,wyart:17,berthier:19}. An explanation stemming from an equivalence with spin glasses would be that those are preasymptotic effects preceding a phase transition at lower temperature, though the nature of these effects and of the low-temperature phase still would remain unknown.\\

Our simulations give evidence that the mean-field model describing appropriately the three-dimensional Heisenberg spin glass is not the SK model, nor its extension to vector spins~\cite{elderfield:82,cragg:82},
which have infinite steps of replica symmetry breaking. Instead, it is most likely that it be the (mixed~\cite{folena:19}) $p$-spin model, with only one step of replica symmetry breaking, that has a phenomenology similar to the Heisenberg spin glass. This observation suggests that the high-dimensional (mean field) limit of low-dimensional spin glasses does not share their universality class.
This picture is further sustained by earlier work, showing that the chiral overlap distribution in the Heisenberg spin glasses is 1-RSB~\cite{hukushima:05,takahashi:15}.\\

On a last note, we remark that the dynamical crossover we measured is different in the spin and chiral glass sectors~\cite{picco:05,kawamura:10}. This opens a new scenario for the spin-chirality decoupling debate~\cite{kawamura:92}, regarding the putable presence of two critical temperatures in the Heisenberg spin glass: the spin-chirality decoupling could be attributed to a different dynamical crossover in the two critical channels.

\section{Acknowledgments}
We thank Federico Ricci-Tersenghi for many valuable discussions.
We thank Luis Antonio Fern\'andez for suggesting the use of the Successive Overrelaxation algorithm. We also thank Ada Altieri, Silvio Franz and Harukuni Miyazaki for interesting discussions.
This work was funded through Grant No. FIS2015-65078-C2-1-P, jointly funded by MINECO (Spain) and FEDER (European Union).
M.B.-J. was supported by the Simons Foundation for
the collaboration “Cracking the Glass Problem” (No. 454951 to D.R. Reichman). 


\bibliographystyle{unsrt}
\bibliography{marco}

\end{document}